%% file: main.tex
\documentclass[runningheads]{llncs}

\input{config}
\usepackage{appendix}

\usepackage{paralist}
\usepackage{floatrow}
\usepackage{mathtools}
\input{mymacros}

\newcommand{\nbnote}[3]{
  \fcolorbox{gray}{yellow}{\bfseries\sffamily\scriptsize#1}
            {\color{#2}
              \sffamily\small$\blacktriangleright$\textit{#3}$\blacktriangleleft$}
}
\newcommand{\mario}[1]{\nbnote{Mario}{red}{#1}}
\renewcommand{\mario}[1]{}
\newcommand{\ion}[1]{\nbnote{Ion}{blue}{#1}} %
\renewcommand{\ion}[1]{}%

\usepackage{tikz}

\newcommand\myheight{1.2ex} % cup and cap height
\newcommand\mywidth {1.3ex} % cup and cap width
\newcommand\mylinew {0.17ex} % cup and cap line width
\newcommand\spaceba {0.01ex} % space before and after
\newcommand\mystart {-0.05ex} % space before and after
\newcommand{\mysymbol}[1]
{
  \hspace{\spaceba}
  \tikz[line width=\mylinew,line cap=round,rotate=#1]
    \draw[transform canvas={yshift=\mystart}] (0,\myheight) -- (0,0.5*\mywidth) arc (-180:0:0.5*\mywidth) -- (\mywidth,\myheight);
  \hspace{\spaceba}
}
\newcommand{\mycup}{\mysymbol{0}}

\begin{document}

\title{Unfolding Iterators: Specification and Verification of
  Higher-Order Iterators in \ocaml}
\titlerunning{Unfolding Iterators}
\author{ Ion Chirica \and Mário Pereira }
\authorrunning{I. Chirica, M. Pereira} \institute{NOVA LINCS, Nova
  School of Science and Technology, Portugal
  }
\maketitle
\begin{abstract}

Albeit being a central notion of every programming language, formally
and modularly reasoning about iteration proves itself to be a
non-trivial feat, specially in the context of higher-order
iteration. In this paper, we present a generic approach to the
specification and deductive verification of higher-order iterators,
written in the \ocaml language. Our methodology follows two key
principles: first, the usage of the \gospellang specification language
to describe the general behaviour of any iteration schema; second, the
usage of the \cameleer framework to deductively verify that every
iteration client is correct with respect to its logical
specification. To validate our approach we develop a set of verified
case studies, ranging from classic list iterators to graph algorithms
implemented in the widely used \textsf{OCamlGraph} library.

  % Deductive Software Verification
  % Formal Specification
  % Higher-Order Iteration
  % OCaml
  % Gospel
  % Cameleer

\end{abstract}

% \ion{escolher um standard sobre cursor, ou é \emph{cursor} ou cursor}
\section{Introduction}
\label{sec:introducao}

Iteration plays a crucial role in every programming language. It
comes in various shapes, be it through cycles, recursive functions
or even higher-order iterators. The usage of an iteration process is
normally associated with the need to repeat a computation,
\emph{e.g.}, process every element in a data structure.
Nonetheless, despite our extensive experience with the use of
iteration, implementing and ensuring that an iterative algorithm is
exempt from errors remains a non-trivial task.

Testing is a very popular approach to analyze a given program.
It is a simple and fast methodology that can be easily incorporated
into traditional development cycles. However, tests are rarely
exhaustive and for that reason, their correctness guarantees are as
limited as their coverage. If we wish to give the highest assurance of
program correctness, we quickly turn to {formal
  methods}~\cite{DBLP:books/daglib/0007501}.
This field uses mathematical tools to reason about program
correctness. In particular, we focus on \emph{deductive software
verification}~\cite{filliatre11sttt}, which translates correctness
properties into logical implications to be proven by a computer. This
technique is also known as \emph{formal software proving}.

In this work we tackle the problem of formally verifying programs that
employ higher-order iterators.  The iteration patterns defined within
the class of higher-order iterators are so common and well established
that we believe a \emph{generic} specification should be able to
formally capture its behaviour.
Our purpose is to establish a
\emph{quasi}-automatic verification methodology, in a way that we
avoid the cumbersome effort of conducting such reasoning in a proof
assistant.
The programs of interest are written in
\ocaml, a language of the ML family, where iteration is naturally
expressed using higher-order functions. While \ocaml allows combining a functional
programming style with imperative mechanisms, notably memory writes,
this turns out to be a challenge for deductive verification tools.
To address these challenges, we present
\begin{inparaenum}[(1)]
\item a generic way to specify high-order iterators where we capture,
in first-order logic, essential patterns that any iterator should
adopt;
\item a proof methodology based on translating a higher-order iterator
into an equivalent program in first-order, utilizing the notion of
cursor, \emph{i.e.}, a step-by-step iterator.
\end{inparaenum}
Our approach is implemented as a set of extensions to the
specification language \gospellang~\cite{ChargueraudFLP19} and
\cameleer~\cite{pereiracav21}, a deductive verification tool for
\ocaml programs. All the examples are available in an
online artifact~\cite{ifm2025}.

\section{Background}
\label{sec:raciter}

\paragraph{Reasoning about iteration.}
% Reasoning about iteration is not novel, first explored in Turing's
% foundational work to model repetitive processes, and later formalized
% by Tony Hoare through axiomatic semantics and the introduction of loop
% invariants.
The effort to formally reason about iteration is not novel, having
been explored through loop invariants~\cite{hoare}, verification
condition generation~\cite{wpc}. More recently, Filliâtre and
Pereira~\cite{nfm16} proposed a modular approach using two predicates
per iterator: \texttt{permitted} (for values produced during
iteration) and \texttt{completed} (for termination), and a sequence of
visited elements. Exemplifying, when
iterating over a sequence \emph{s}, the \gosp{permitted} predicate
ensures that any sequence of visited elements \emph{v} must be a valid
prefix of \emph{s}, formally:
\[
\texttt{permitted}(v, s) \defen ||v|| \leq ||s|| \land
\forall i. 0 \leq i \le ||v|| \implies v[i] = s[i]
\]
On the other hand, the \gosp{complete} predicate then simply states
that the iteration is finished when the length of the visited sequence
\emph{v} equals the length ($||\cdot||$) of \emph{s}, formally
captured with:
\[
\texttt{complete}(v, s) \defen ||v|| = ||s||
\]

Based on these two predicates, we are able to capture any kind of
iteration, be it over a collection or enumeration of results of an
algorithm, be it finite and deterministic or infinite and
non-deterministic~\cite{nfm16}.

\paragraph{Specification and Verification of \ocaml
programs.}\gospellang (Generic Ocaml SPEcification Language) is a
behavioural specification language for \ocaml
interfaces~\cite{ChargueraudFLP19}.  It is a contract-based, strongly
typed language with formal semantics defined employing translation
into Separation Logic~\cite{sl}. The main goal of \gospellang is to
provide a concise and accessible specification language for \ocaml
interfaces, which can be used for various purposes. However, with a
strong emphasis on verification with the use of tools that translate
\gospellang annotated \ocaml into languages understandable by
automated theorem provers. \gospellang specifications are added as
comments beginning with \texttt{@}, \texttt{(*@ ... *)}, at the end of
the function definition.
Conversely, \cameleer, is a tool for deductive verification of \ocaml
programs, conceived with proof automation in
mind~\cite{pereiracav21}. This tool takes \gospellang, additionally
supporting implementation specifications, annotated \ocaml code and
translates it to an equivalent \whyml~\cite{why}, which
can be verified using automated provers like \cvc, \altergo, or
\zthree, or interactive ones such as \coq and \isabelle.

\section{Motivating example}

To motivate our approach and tool choices, we consider summing a
finite, deterministic sequence of integers using two classic patterns:
recursion and loops. Later, we revisit this example using cursors and
higher-order iterators.

From a recursive perspective, the function sums elements from
\gosp{upper} to \gosp{lower}, using \gosp{get} to access elements
(whose complexity we ignore in the context of this
example). Termination is ensured by making \gosp{upper} the decreasing
argument and bounding both \gosp{lower} and \gosp{upper} within
\gosp{s}. We ensure that the result matches the mathematical sum over
that range.
\begin{gospelsmall}
  let rec sum_recursive s lower upper =
    if upper <= lower then 0
    else (get s (upper - 1)) + sum_recursive s lower (upper - 1)
  (*@ r = sum_recursive s lower upper
      variant upper
      requires 0 <= lower <= upper <= *?$||$?*s*?$||$?*
      ensures r = *?$\sum_{\texttt{i=lower}}^{\texttt{upper}} \texttt{s[i]}$?**)
\end{gospelsmall}
Using a cycle, this can be solved by introducing two mutable
variables, \gosp{index} to get the \emph{n-th} element of a sequence,
and \gosp{acc} the accumulator of the sum. In this cycle, we iterate
over the length of \gosp{s} and successively add its elements:
\begin{ocamlsmall}
  let sum_cycle s =
    let index = ref 0 and acc = ref 0 in
    while !index < (length s) do
\end{ocamlsmall}
The loop specification first states that the termination measure is
given by a decrease in the to-be visited elements, followed by the
invariant: \gosp{index} is contained within the limits of the sequence
and the contents of \gosp{acc} holds the current sum of the elements
on \gosp{s}, up until \gosp{!index}, as follows:
\begin{gospelsmall}
    (*@ variant *?$||$?*s*?$||$?* - !index
        invariant 0 <= !index <= *?$||$?*s*?$||$?* /\ !acc = *?$\sum_{\texttt{i=0}}^{\texttt{!index}} \texttt{s[i]}$?* *)
      acc := !acc + (get s !index);
      index := !index + 1;
    done;
    !acc
\end{gospelsmall}
Finally, the specification of \gosp{sum_cycle} is given by the
post-condition that the result, \gosp{r}, of calling this function
equals the logical definition of a sum:
\begin{gospelsmall}
  (*@ r = sum_cycle s
      ensures r = *?$\sum_{\texttt{i=0}}^{||\texttt{s}||} \texttt{s[i]}$?**)
\end{gospelsmall}
This specification we present here is a very high-level one, however,
there is a definition within \gospellang's Standard Library for the
summation operator. When fed to \cameleer, the above specifications
and \ocaml implementations, will be translated into a \whyml program
for which \whythree generates a total of 9 Verification Conditions
(VCs), 4 for the recursive definition and 5 for the cycle based one,
all of them quickly dispatched by the \altergo SMT solver.

\section{Revisiting modular specification for first order iteration}
\label{sec:spec}
In this section we present a refinement of the existing \whyml
cursor specification. This refinement is a key part of our translation
schemas (c.f. \Cref{sec:translation}), where we resort to \whythree's
module system to achieve a generic specification for cursors.

Cursors are defined using two functions, \texttt{has\_next()} and
\texttt{next()}, testing whether the iteration has completed, and
returning the next element in the iteration, respectively. Building on
the ideas of Filli\^atre and Pereira~\cite{nfm16}, we propose an
extension to their cursors in \whyml. This extension, unable at the time due to
limitations in \whythree to express type invariants, embeds the
predicates \gosp{permitted} and \gosp{complete} within the type of the
cursor, as follows:
\begin{whylang}
  type cursor 'a = abstract {
    mutable visited   : seq 'a;
            permitted : seq 'a -> bool;
            complete  : seq 'a -> bool;
  } invariant { permitted visited }
\end{whylang}
The sequence of visited elements, the first field, \gosp{visited},
holds the sequence of polymorphic elements \gosp{visited}, and the
other two are predicates, parameterized with a sequence of polymorphic
elements. By embedding these predicates within the \texttt{cursor}
type, we generalize the iteration to any collection. Additionally,
this allows us to decorate this type definition with a type invariant, stating
that: at any point before a function call, possibly invalidated in
between, but fully restored by the function's exit, the sequence of
visited elements, \gosp{visited}, is \texttt{permitted}. This type is
\gosp{abstract}, that is, its fields are visible in the specification
while being opaque in the implementation.

This refinement of a cursor and the functions \gosp{next} and
\gosp{has_next} are fully generic, irrespective of which iteration we
desire to express. Notwithstanding, we lack a function that creates a
cursor, \gosp{create}. Fundamentally, this function differs from
cursor to cursor, and thus unable to be generic. For that reason, the
definition of this function is delegated to the cursor client,
allowing for iterations over diverse data structures or the
specification of custom iteration patterns.

Coming back to our recurring theme of iterations over sequences, we
can finish the definition of such a cursor by instantiating its
fields, like so:
\begin{whylang}
  val create (s: seq 'a) : (c : cursor 'a)
    ensures { result.visited   = empty }
    ensures { result.permitted = (fun v -> *?$||\texttt{v}||$?* <= *?$||\texttt{s}||$?* /\
                                  forall i. 0 <= i < *?$||\texttt{v}||$?* -> v[i] = s[i]) }
    ensures { result.complete  = (fun v -> *?$||\texttt{v}||$?* = *?$||\texttt{s}||$?*) }
\end{whylang}
The first post-condition is natural, at creation we ensure a cursor
with an empty sequence of visited elements. Afterwards, we give body
to the predicates \texttt{permitted}, with the definition of
\texttt{v} being a prefix of the collection to be iterated, and the
\texttt{complete} given by the comparison of both lengths.
The direct conclusion is that by instantiating the pair
\texttt{permitted}/\texttt{complete} at the cursor creation, we claim
the same result as the work of Filli\^atre and Pereira \cite{nfm16}.

Having defined a cursor over sequences, let us now see how a client
might use this construct to sum all the elements of a sequence of integers:

\begin{whylang}
  let sum_cursor (s: int seq) : int
*?\codemark{A}?* ensures { result = sum (fun i -> s[i]) 0 (length s) } =
    let acc = ref 0 in
    let c = create s in
*?\codemark{B}?* while has_next c do
        variant { *?$||\texttt{s}||$?* - *?$||\texttt{c.visited}||$?*  }
*?\codemark{C}?*   invariant { !acc = sum (fun i -> c.visited[i]) 0 *?$||\texttt{c.visited}||$?* }
*?\codemark{D}?*   let x = next c in
      acc := !acc + x;
    done;
    !acc *?\begin{tikzpicture}[main/.style = {draw, circle}, overlay, remember picture,el/.style = {inner sep=2pt, align=left, sloped},every label/.append style = {font=\footnotesize}]
  	\draw[dashed, line width=0.25mm, dash pattern=on 2pt off 2pt] (B.south) -- (C.north);
\end{tikzpicture}?*
\end{whylang}
Its specification can be given in the following way: in \codemarkt{A}
we ensure that the result is equal to the mathematical
sum\footnote{The function \gosp{sum f l u}, defined as the sum
$\sum_{\texttt{i=l}}^{\texttt{u}} \texttt{f(i)}$, is included in the
\whythree Standard
Library. \url{https://www.why3.org/stdlib/int.html}} of all the
elements of \gosp{s}; from \codemarkt{B} to \codemarkt{C} we find the
test of termination, the cycle variant and invariant, given by the
accumulator being the sum of all the visited elements; finally, in
\codemarkt{D} we produce an element, \gosp{x}, and the subsequently
update to the accumulator with the sum of \gosp{acc} and \gosp{x}.

If we instantiate \whythree on \gosp{sum_cursor} and the presented
specification, we end up with 6 VCs, three
of them concerned with the loop invariants, two with the preconditions
of the \gosp{next} and \gosp{has_next} and the last with the
post-condition of \gosp{sum_cursor}. All were fully dispatched by
\altergo.

\section{Higher-order iteration specification}
\label{sec:hoi-spec}

Computing the sum of all elements in a sequence naturally aligns with
a well-known higher-order iteration pattern: a so called fold
function. This is ubiquitous in mainstream languages such as \ocaml,
\textsf{Haskell}, or \textsf{Scala}. In this section, we formalize
such a computation by considering a parameterized OCaml module,
\gosp{Sequences}, which captures the essence of this higher-order
abstraction.

This module is parameterized over another module \gosp{S},
which specifies a polymorphic sequence type \gosp{seq}\footnotemark{}
and two higher-order functions: \gosp{fold} and \gosp{iter}. The
function \gosp{fold} takes three arguments: a binary function of type
\gosp{'a -> 'b -> 'a}, an initial accumulator value of type \gosp{'a},
and a sequence of type \gosp{'b seq}, returning a final accumulated
value of type \gosp{'a}. The second function, \gosp{iter}, applies a
function of type \gosp{'a -> unit} to each element of a given
sequence, solely for its side effects.

In the \gosp{Sequences} module, we define \gosp{sum_fold}, which
computes the sum of a sequence using \gosp{fold} with the binary sum
as the consumer function and
initial value \gosp{0}. We also define \gosp{stack_of_seq}, which
builds a stack from a sequence using \gosp{iter}, where the consumer
function pushes elements onto the stack in-place.

\footnotetext{This type denotes a sequence: an ordered collection of
elements in which repetitions are permitted, \emph{e.g.}, lists or
arrays. It should not be mistaken with \ocaml's \gosp{Seq.seq} type.}

\begin{ocamlsmall}
  module Sequences (S : sig
    type 'a seq
    val fold : ('a -> 'b -> 'a) -> 'a -> 'b seq -> 'a
    val iter : ('a -> unit) -> 'a seq -> unit
  end) = struct
    let sum_fold (s : int S.seq) = S.fold (fun acc x -> acc + x) 0 s

    let stack_of_seq s =
      let stack = Stack.create () in
      S.iter (fun x -> Stack.push x stack) s; stack
  end
\end{ocamlsmall}

% let sum_fold_mut (s : 'a Seq.seq) =
%   let acc = ref 0 in
%   ignore (Seq.fold (fun _ x -> acc := !acc + x) () s);
%   ! acc

In the previous section, we presented a generic approach to iteration
through the use of cursors, focusing exclusively on the definition of
the \texttt{create} function. We observed that iteration is a two-fold
task: first, abstracting what it means to iterate over an arbitrary
collection; and second, defining the concrete semantics of the
iteration process itself. The former is captured by an \ocaml
interface, in this context, via the definition of the \gosp{fold}
function, while the latter is tied to the implementation, specifically
when \gosp{fold} is invoked.

In the remaining of this section, we demonstrate how such constructs can be formally
specified using \gospellang. We then present our extensions to the
language, together with their translation into \whyml via
\cameleer. Our focus lies in providing specification support
for the most commonly used iteration patterns, namely \texttt{iter},
\texttt{fold}, \texttt{filter}, and \texttt{map}. Furthermore, we
argue that supporting \texttt{fold} alone would be theoretically
sufficient, as the remaining patterns can be viewed as
degenerate forms of the fold abstraction.

\subsection{Interface overview}
\label{sec:int-overview}

Considering the previous example, let us explain how we can define an
iteration over a sequence, using our extension to the \gospellang
language, and in terms of \texttt{permitted} and \texttt{complete}.
\gospellang specifications of functions start by introducing
arbitrary, and user-selected, names of the function's result, \gosp{r}
and its arguments, \gosp{func}, \gosp{acc} and \gosp{col}:
\begin{gospelsmall}
  val fold: ('a -> 'b -> 'a) -> 'a -> 'b seq -> 'a
  (*@ r = fold func acc col
\end{gospelsmall}
Our extensions to the language begin by specifying which iteration
pattern we wish to define, and dealing with a fold, we indicate this
using the keyword:
\begin{gospelonly}
    folds
\end{gospelonly}
After that, we instantiate the predicates \texttt{permitted} and
\texttt{complete}, both of which take the sequence of visited elements
as an argument. To refer to the collection being iterated over within
these predicates, one utilizes the keyword \gosp{collection}:
\begin{gospelonly}
    ~permitted:(fun v -> *?$||\texttt{v}||$?* <= *?$||\texttt{collection}||$?* /\
                forall i. 0 <= i < *?$||\texttt{v}||$?* -> v[i] = (collection)[i])
    ~complete:(fun v -> *?$||\texttt{v}||$?* = *?$||\texttt{collection}||$?*)
\end{gospelonly}
At this point, we have to be explicit about some typing information:
the type of the structure we are iterating over, \gosp{structure}, the
type of elements that the cursor produces, \gosp{elt}, and as
auxiliary information, what is the name of the \gosp{accumulator} in
the header of the specification:
\begin{gospelonly}
    with structure = ('b seq), elt = 'b, accumulator = acc *)
\end{gospelonly}
% This fairly simplistic specification lets us capture everything we
% need to define a cursor over a polymorphic sequence.

The reader may have noticed that, although an argument of \gosp{fold},
we make no mention of the consumer function, \gosp{func}.
Part of turning a higher-order iterator into a first-order cursor
client is delaying acknowledging its existence until it has been
instantiated.
In fact, our specification of \gosp{fold} abstracts away three
fundamental aspects of the consumer function:
\begin{inparaenum}[(1)]
\item the only side-effects performed by the higher-order iteration
  are those of the consumer function.
\item the only exceptions, potentially, raised by the higher-order
  iteration are those raised by the consumer function.
\item we state the consumer function re-establishes the user-supplied
  invariant, step-by-step, for each new enumerated element.
\end{inparaenum}

\subsection{Implementation overview}
\label{sec:impl-overview}

Having covered how one might define a generic cursor in an interface,
let us now see how one might go onto specifying its behaviour. This is
only possible to do when one calls \gosp{fold}, during the
implementation. Recalling the function \gosp{sum}:

\begin{ocamlsmall}
  let sum (s: S.seq) = S.fold (fun a x -> a + x) 0 s
\end{ocamlsmall}
This function employs the higher-order iteration function
\gosp{fold}. However, its implementation is opaque to the
user. They do not know how the iteration takes place, but they expect
the consumer function to be applied to every element of \gosp{s},
with an initial accumulator 0. We argue that by hiding how the
iteration takes place in logic, we only ask the user to specify the
essentials.
Specifying its semantics first starts by stating the kind of iteration
pattern they expect, and seeing as we are dealing with a \gosp{fold}, we
start with the keyword:
\begin{gospelonly}
  (*@ folds
\end{gospelonly}
Next, and in no particular order, they indicate the collection being
iterated over:
\begin{gospelonly}
      ~collection:s
\end{gospelonly}
The termination measure, in the case of finite and deterministic
enumerations, follows a pattern: the number of to-be-visited elements
converging to zero:
\begin{gospelonly}
      ~convergence:(fun c v -> *?$||\texttt{c}||$?* - *?$||\texttt{v}||$?*)
\end{gospelonly}
And lastly, the iteration invariant:
\begin{gospelonly}
      ~inv:(fun a v -> a = sum (fun i -> v[i]) 0 *?$||\texttt{v}||$?*) *)
\end{gospelonly}
This new syntax may have introduced a lot of new ideas at once. Let us
make them clear. As we have also seen in the interface, we have to
identify what iteration pattern we wish to express, through the
keyword \gosp{folds}, which helps to decide what first-order client
schema to generate during translation. All the other parametric fields
are \gospellang logical terms. Note how the parameter
\gosp{convergence} takes two arguments, the collection that is being
iterated and the sequence of visited elements. The \gosp{inv}, in the
case of a {fold}, which introduced the notion of an accumulator, is
also parameterized with two arguments, the current value of the
accumulator and the sequence of visited elements. In both cases, the
arguments of \gosp{convergence} and \gosp{inv}, are automatically
applied during translation.

\section{Translation-based approach to iteration semantics}
\label{sec:translation}

% In previous sections, we explored how to tell an iteration tale by
% defining generic cursors and performing first-order iteration using
% their clients in \whyml. We then showed how to specify higher-order
% iteration in \gospellang.

In this section we present our extensions to \cameleer through
translation schemas. Though we implemented four schemas, one for each
iteration pattern, for presentation purposes, we will only focus on
the \texttt{fold}. Once again, we argue that the other patterns
(\texttt{iter}, \texttt{filter}, and \texttt{map}) are degenerate
cases of \texttt{fold}. Specifically, \texttt{iter} corresponds to a
\texttt{fold} where the accumulator has type \texttt{unit}, and
\texttt{map} can be expressed as a \texttt{fold} that builds a new
collection by applying a function to each element and adding the
result. Likewise, \texttt{filter} can be implemented as a
\texttt{fold} that adds elements to the new collection only if they
satisfy a predicate.

\subsection{Translating iteration declarations}

The first translation schema is concerned with a higher-order function
declaration. These declarations can be found in \ocaml interface files
or as arguments of a functor. The general specification form for a
fold is the following:
\begin{gospelsmall}
  val fold: ('a -> 'b -> 'a) -> 'a -> 'b t -> 'a
  (*@ r = fold func acc col
      folds ~permitted:(fun v -> *?\textit{term}$_p$?*) ~complete:(fun v -> *?\textit{term}$_c$?*)
      with structure = *?$\tau_s$?*, elt = *?$\tau_e$?*, accumulator = acc *)
\end{gospelsmall}
As we have seen, \gosp{permitted} and \gosp{complete} are defined using
\gospellang logical terms. The types \gosp{structure} and \gosp{elt}
are also part of the specification, and the \gosp{accumulator} is
identified by its name in the header.

With this context in place, we are now in a position to translate the
specification into the creation of a cursor. In \cameleer, this
translation replaces the original function declaration with a
\texttt{scope}, that is, a namespace~\cite{genericity}.
The translation schema follows the general shape:
\begin{whylang}
  scope Fold
    use seq.Seq
    clone export cursor.CursorLib
    val create (collection: *?$\tau_{s}$?*) : cursor *?$\tau_{e}$?*
      ensures { result.visited = empty }
      ensures { result.permitted = (fun v -> *?\textit{term}$_p$?*) }
      ensures { result.complete  = (fun v -> *?\textit{term}$_c$?*) }
  end
\end{whylang}
Within the scope, we clone the \gosp{CursorLib} module, containing the
developments from ~\Cref{sec:spec}, to expose the generics of a
cursor, and define the \gosp{create} function by extracting and
translating the types, $\tau_{s}$ and $\tau_{e}$, and terms,
\texttt{\textit{term}}$_p$ and \texttt{\textit{term}}$_c$.

\subsection{Translating an iteration call}
When it comes to iteration clients, we extend \cameleer to translate
the specification below into a first-order iteration using a cursor
client:
\begin{gospelsmall}
  fold (fun a e -> ...) col x*?$_{\texttt{0}}$?*
  (*@ folds ~inv: (fun a v -> *?\textit{term}$_i$?*) ~collection: *?\textit{term}$_{c'}$?*
      ~convergence: (fun c v -> *?\textit{term}$_v$?*) *)
\end{gospelsmall}
Our translation schema starts by initializing the accumulator,
\gosp{acc}. We are able to get the value, \texttt{x}$_{\texttt{0}}$,
as we have identified in the interface specification it in the list of arguments.  We then
follow by creating a cursor over the collection \texttt{\textit{term}}$_{c'}$:
\begin{whylang}
  let acc = ref x*?$_{\texttt{0}}$?* in
  let cursor = Fold.create *?\textit{term}$_{c'}$?* in
\end{whylang}
Here we find the while-loop, conditioned by the \gosp{has_next}
function. Note that if this loop ends, then when it does,
the cursor predicate \gosp{complete} holds:
\begin{whylang}
  while Fold.has_next cursor do
\end{whylang}
The variant of the loop is \texttt{\textit{term}}$_{v}$, where we apply
\texttt{\textit{term}}$_{c'}$, the collection that is being iterated, and the
sequence of visited elements. On the other hand, the invariant is
\texttt{\textit{term}}$_{i}$, where we apply the accumulator and the visited sequence:
\begin{whylang}
      variant { (fun c v -> *?\textit{term}$_{v}$?*) *?\textit{term}$_{c'}$?* cursor.visited }
    invariant { (fun a v -> *?\textit{term}$_i$\hspace{0.1em}?*) !acc cursor.visited }
\end{whylang}
We produce the next element in the iteration and store it in
\gosp{x}. Followed by the application of the consumer function to the
accumulator and \gosp{x}:
\begin{whylang}
    let x = Fold.next cursor in
    acc := (fun a e -> ...) !acc x;
  done;
\end{whylang}
Finally, we return the contents of the \gosp{accumulator}:
\begin{whylang}
  !acc
\end{whylang}
This concludes our translation schema, which makes higher-order
specifications amenable to automated verification by translating them
into a first-order form.
% \subsection{Argument of Degeneracy}

% We have defined translation schemas for four of the most common
% high-order iterator patterns: \gosp{iter}, \gosp{map}, \gosp{fold} and
% \gosp{filter}. We wish to make an argument that we are able to express
% all the other by only using a \gosp{fold}.

\subsection{Capturing nested iterations}

A natural question is: what about nested iteration, where the consumer
function itself uses a higher-order iterator?
By analogy with classical reasoning, nested loops require the inner
one to preserve the invariant of the outer. Consider the following
example, adapted to our translation:
\begin{whylang}
  while has_next cursor do
   *?\codemark{A}?*invariant { I !a v } ...
    while has_next cursor' do
     *?\codemark{B}?*invariant { I' !a' v' } ...
\end{whylang}
We aim to preserve the invariant \codemarkt{A}, parameterized by the
current accumulator~\gosp{!a} and visited elements \gosp{v}, inside
the inner loop \codemarkt{B}. We track the history of these arguments:
for \gosp{fold}, \gosp{map}, and \gosp{filter}, both accumulator and
visited elements; for \gosp{iter}, only the visited elements. During
translation, we propagate these arguments to inner invariants, which
must now account for:

\begin{whylang}
 *?\codemark{A}?*invariant { I !a v }
 *?\codemark{B}?*invariant { I' !a' v' !a v }
\end{whylang}
More generally, we append ($\oplus$) the arguments from all levels
above the current one, $l$, while accounting for partial applications,
represented by $\overline{{\partial{}}}$ as a
sequence of partially applied arguments:
\begin{whylang}
    *?\hspace{0.05em}?*invariant { I *?$\overline{{\partial{}}}$?* (*?$\mathop{\oplus}\limits_{i=0}^{l}$?* (!a*?$_i$?*)? (v*?$_i$?*)) }
\end{whylang}

\section{Coping with mutability}

We have promised that our methodology is also capable of capturing
iterations where the consumer function performs side-effects,
\emph{e.g.}, writes to a reference. Let us present an example of a
function that produces a stack from a sequence. This function employs
the \gosp{iter} higher-order iterator to traverse a sequence and
systematically push elements to the stack, modifying the stack
in-place.  Recalling the \gosp{iter} function, previously presented in
\Cref{sec:hoi-spec}, we can decorate it with a specification in the
following way:
\begin{gospelsmall}
  val iter: ('a -> unit) -> 'a seq -> unit
  (*@ r = iter func col
    iters ~permitted:(fun v -> *?$||\texttt{v}||$?* <= *?$||\texttt{collection}||$?* /\
                      forall i. 0 <= i < *?$||\texttt{v}||$?* -> v[i] = (collection)[i])
          ~complete:(fun v -> *?$||\texttt{v}||$?* = *?$||\texttt{collection}||$?*)
    with structure = ('a seq), elt = 'a *)
\end{gospelsmall}
This specification is very similar to that of \gosp{fold}, with the
primary differences being the use of the \gosp{iters} keyword and the
absence of an explicit accumulator. This is expected, as both
iterators operate over the same data structure and in the same
direction, resulting in identical definitions for \gosp{permitted} and
\gosp{complete}.

Consider the \gosp{stack_of_seq} function: it begins by creating an
empty stack, modeled logically as a list, and then uses \gosp{iter} to
traverse the sequence \gosp{s}, pushing each element onto the
stack. Once iteration completes, the stack is returned.
\begin{gospelsmall}
  let stack_of_seq s =
    let stack = Stack.create () in
    S.iter (fun x -> Stack.push x stack) s
    (*@ iters ~inv:(fun v -> reverse stack = s[..*?$||$?*v*?$||$?*])
        ~collection:s ~convergence:(fun c v -> *?$||$?*c*?$||$?* - *?$||$?*v*?$||$?*) *); stack
  (*@ r = stack_of_seq s
      ensures reverse r = s *)
\end{gospelsmall}
The iteration invariant states that the reverse of the stack,
reflecting insertion order, equals the visited prefix, \gosp{s[..i]}
is the prefix of \gosp{s} until \gosp{i}. The variant is the
decreasing number of elements left to visit.  The postcondition
follows from termination: once \gosp{has_next} returns false, the loop
exits, ensuring that \gosp{complete} holds. By the invariant, the
accumulator then contains the full reverse of \gosp{s}.
Despite the side effects, Alt-Ergo discharges the VCs, in under 3
seconds.

% Note that in this case, we rely on the fact that
% \gosp{Stack} is implemented in both \ocaml's and \whyml's standard
% libraries. During translation, with \cameleer, and verification,
% operations such as \gosp{push} are interpreted according to \whyml's
% definition. In \whyml, the stack is modeled as an abstract type with a
% mutable field \texttt{elts} representing the list of elements.

% \begin{whylang}
%   type t 'a = abstract { mutable elts: list 'a }
%   val push (x: 'a) (s: t 'a) : unit writes {s}
%     ensures { s.elts = Cons x (old s.elts) }
% \end{whylang}

% This function modifies the stack in place: \texttt{push} prepends the
% element \texttt{x}, preserving order.

\section{Case study: \textsf{OCamlGraph}}
\label{sec:cases}

In this section, we present our main case study, proving the
correctness of modules taken from \textsf{OCamlGraph}, a generic
library with a wide range of graph structures and
algorithms~\cite{ocamlgraph}. Widely used across \textsf{OPAM}
packages, it relies on higher-order iterators, making it a suitable
candidate to validate our methodology.

A particular module within \ocamlgraph we aimed to prove correct is the
module \gosp{Oper}\footnotemark{}. This module provides common operations over
graphs, such as, the intersection and union of two graphs, and the
complement and mirror of a graph. In the subsequent sections, we present a
logical model of a graph, the specification of iterations over graphs,
and illustrate a specification and application of our methodology to the union of
two graphs.

\footnotetext{\url{https://github.com/backtracking/ocamlgraph/blob/master/src/oper.ml}}

\subsection{Logical model of a graph}

The mathematical definition of a graph is through sets, and for our
purposes these will be finite. It is modeled using a domain, the set
of vertices that make up a graph, and a function that takes a vertex
and returns a set vertices, these will be our successors of a
vertex. In other words, the set of edges in the graph.

The module \gosp{Oper} takes as an argument another module, \gosp{G}, that
represents our graph.
The module \gosp{Oper} starts by defining the type of a vertex and a graph:
\begin{ocamlsmall}
  module Oper (G : sig
    type vt (* arbitrary vertex type *)
    type gt (* arbitrary graph type *)
\end{ocamlsmall}
The type graph is annotated with a logical model. It is made up of a
domain \gosp{dom} and the map \gosp{suc}. Additionally, we give this
type an invariant stating that the graph's domain is closed under the
map \gosp{suc}, and for every element that is not in the domain, its
set of successors is empty.
\begin{gospelonly}
    (*@ model dom: vt fset
        model suc: vt -> vt fset
        invariant (forall v1, v2. v1 *?$\in$?* dom /\ v2 *?$\in$?* (suc v1) -> v2 *?$\in$?* dom) /\
                  (forall v1. not_(v1 *?$\in$?* dom) -> (suc v1) == *?$\emptyset$?*) *)
\end{gospelonly}
An empty graph is not very useful, so we define two functions to build
its domain and successors: \gosp{add_vertex} and \gosp{add_edge}. The
first takes a graph and a vertex and returns a new graph with the vertex
added to the domain and the successor map remains unchanged:
\begin{gospelsmall}
    val add_vertex : gt -> vt -> gt
    (*@ g' = add_vertex g v
        ensures g'.dom = add v (g.dom) /\ forall v. g'.suc v = g.suc v *)
\end{gospelsmall}
The second function adds an edge from \gosp{v} to \gosp{v'}, returning a
graph where the domain is unchanged, and the successor map updates \gosp{v}
with the addition of \gosp{v'}:
\begin{gospelsmall}
    val add_edge : gt -> vt -> vt -> gt
    (*@ g' = add_edge g v v'
        ensures g'.dom = g.dom /\ g'.suc = g.suc[v <- add v' (g.suc v)] *)
\end{gospelsmall}
Finally, let us also specify a function that yields a copy of a graph,
\gosp{copy}:
\begin{gospelsmall}
    val copy: gt -> gt
    (*@ g' = copy g
        ensures g.dom = g'.dom /\ forall v. g.suc v = g'.suc v *)
\end{gospelsmall}
These functions cover the necessary ground to build graphs, expanding
its domain set or set of successor. We make use of these functions later on.

\subsection{Higher-order iterators over graphs}

We have covered how to build graphs, but we are yet to mention how to
iterate over one. The module \gosp{Oper} makes use of two functions,
\gosp{fold_vertex}, \gosp{fold_succ}, to iterate over graphs, more
precisely, over finite sets. Let us see how we can specify an
iteration over vertices with the first function:
\begin{ocamlsmall}
    val fold_vertex : (vt -> 'a -> 'a) -> gt -> 'a -> 'a
\end{ocamlsmall}
In order to produce elements, the visited set has to be a subset
($\subseteq$) of the graph's domain; and have to be all distinct as a
sequence. The iteration is complete when we have visited all the
elements in the graph's domain. From the function's arguments we
conclude that we are iterating over a graph and producing vertices.
\begin{gospelonly}
    (*@ r = fold_vertex func graph acc
      folds ~complete:(fun v -> v = collection.dom)
            ~permitted:(fun v -> v *?$\subseteq$?* collection.dom /\ distinct v)
      with structure = gt, elt = vt, accumulator = acc *)
\end{gospelonly}
The second function that iterates a graph is \gosp{fold_succ}, taking
as an argument a graph and a vertex, while iterating over its successors. Its
signature is as follows:
\begin{ocamlsmall}
    val fold_succ : (gt -> 'a -> 'a) -> 'a -> gt -> vt -> 'a
\end{ocamlsmall}
This iteration is also over sets, with the structure of iteration
being a pair \mbox{$($\gosp{gt}$,$ \gosp{vt}$)$}, as it does not make much
sense to iterate over a vertex without a graph nor over successors of
which we do not know their origin. Both \gosp{permitted} and
\gosp{complete} are in regard to the successors of
\gosp{s} in \gosp{g}. This iteration produces vertices.
\begin{gospelonly}
    (*@ r = fold_succ func acc pair
      folds ~complete:(fun v -> let (g, s) = collection in *?$||$?*v*?$||$?* = *?$||$?*g.suc s*?$||$?*)
            ~permitted:(fun v -> let (g, s) = collection in
                                 v *?$\subseteq$?* (g.suc s) /\ distinct v)
      with structure = (gt * vt), elt = vt, accumulator = acc *)
\end{gospelonly}
We are less precise in the \gosp{complete} predicate: we only say that
they have to be equal in their cardinality ($||\cdot||$) but not in
the enumeration of their elements. This is also correct, from
classical set theory follows that if a set \gosp{s1} is a subset of
\gosp{s2} and their cardinality is the same, then they are the same
set.

\subsection{Union}
The union of two graphs, \gosp{g1} and \gosp{g2}, produces a new graph
that combines their vertex sets and successor functions. This is done
by iterating over each vertex \gosp{s} in \gosp{g1} and adding
\gosp{s} and its successors to a copy of \gosp{g2}, used as an
accumulator. The result satisfies the postcondition of representing
the full union of both graphs.

\begin{gospelsmall}
  let union g1 g2 =
    fold_vertex
      (fun g v ->
        fold_succ (fun e g -> add_edge g v e) (add_vertex g v) g1 v
        (*@ folds ~inv:(union_inner g1 g2 v) ~collection:(g1,v)
           ~convergence:(fun (g, s) v -> *?$||$?*g.suc s*?$||$?* - *?$||$?*v*?$||$?*) *)) g1 (copy g2)
    (*@ folds ~inv:(union_outer g1 g2) ~collection:g1
       ~convergence:(fun g v -> *?$||$?*g.dom*?$||$?* - *?$||$?*v*?$||$?*) *)
  (*@ gr = union g1 g2
      ensures gr.dom = g1.dom *?$\mycup$?* g2.dom
      ensures forall src. (gr.suc src) = (g1.suc src) *?$\mycup$?* (g2.suc src) *)
\end{gospelsmall}
Since we are dealing with two higher-order iterators, each requires an
invariant. The outer one, \gosp{union_outer}, applies to
\gosp{fold_vertex}.  It states that the accumulator's domain is the
union of \gosp{g2}'s domain and the visited vertices of \gosp{g1}; for
visited vertices \gosp{v}, their successors in \gosp{acc} are the
union from \gosp{g1} and \gosp{g2}; and for unvisited vertices,
successors match those in \gosp{g2}.  As the information automatically
applied during translation is insufficient, we manually partially
apply the graphs to both invariants, and additionally the vertex being
iterated to the innermost.
\begin{gospelonly}
  (*@ predicate union_outer (g1 g2: gt)
     (* outer iteration *)  (visited: vt seq) (acc: gt) =
   *?\hspace{0.3em}?* (acc.dom = visited *?$\mycup$?* g2.dom)
  /\ (forall v. v *?$\in$?* visited -> acc.suc v = (g1.suc v) *?$\mycup$?* (g2.suc v))
  /\ (forall v. v *?$\in$?* (acc.dom *?$\setminus$?* visited) -> acc.suc v = g2.suc v) *)
\end{gospelonly}
The inner invariant, \gosp{union_inner}, is more subtle. We begin by
stating that the accumulator graph's domain is the union of the
visited vertices and that of \gosp{g2}:
\begin{gospelonly}
  (*@ predicate union_inner (g1 g2: gt) (src: vt)
     (* inner iteration *)  (visited': vt seq) (acc': gt)
     (* outer iteration *)  (visited: vt seq) (acc: gt)  =
     *?\hspace{0.1em}?*(acc'.dom = visited *?$\mycup$?* g2.dom)
\end{gospelonly}
Successors of unvisited vertices are still those from \gosp{g2}:
\begin{gospelnc}
   /\ (forall v. v *?$\in$?* (acc'.dom *?$\setminus$?* visited) -> acc'.suc v = g2.suc v)
\end{gospelnc}
For visited vertices other than the one currently being processed,
its successors in the accumulator are the union from both graphs:
\begin{gospelnc}
   /\ (forall v. v *?$\in$?* visited /\ v <> src -> acc'.suc v = (g1.suc v) *?$\mycup$?* (g2.suc v))
\end{gospelnc}
Finally, the current vertex \gosp{src} accumulates successors incrementally:
\begin{gospelnc}
   /\ (acc'.suc src = visited' *?$\mycup$?* (g2.suc src)) *)
\end{gospelnc}
%
% . This may not hold in
% general, depending on the structure of the iterators and data
% structures.
Note that we did not use the outer accumulator, since \gosp{acc'} is
already an updated version of \gosp{acc}. \altergo dispatches all
the generated VCs of \gosp{union}.

\subsection{Summary of verified case studies}

In this section we summarize a collection of case studies, and
~\Cref{tab:summary} summarizes some statistics. The examples defined
with an \gosp{iter} all have side-effects. We note that the user never
had to reason about the generated \whyml code, and all the
verification conditions were automatically dispatched. The case-studies and
extensions to the tools are available in an online
artifact~\cite{ifm2025}.

\begin{table}[H]
  \rowcolors{0}{white}{white}
  \footnotesize
  \centering
  \begin{tabular}
    {|>{\raggedright}m{.32\textwidth}|>
    {\centering}m{.08075\textwidth}|>
    {\centering}m{.15\textwidth}|c|>
    {\centering}m{.18\textwidth}|c|}
    \hline
    Iteration client & \# VCs & LoC / LoS & Time (s) & Iteration
    & Effects \\\hline
    \rowcolor{thegray}
    Sequences & 18     & 10 / 12 &  & &  \\
    \hspace{1em} \texttt{sum\_seq} & 6     & 1  / 2 & 10.45 &
    \texttt{fold} & - \\
    \hspace{1em} \texttt{stack\_of\_seq} & 1     & 3  / 2 & 3.01  &
    \texttt{iter} & \check \\
    \hspace{1em} \texttt{queue\_of\_seq} & 1     & 3  / 2 & 1.13 &
    \texttt{iter} & \check\\
    % \hspace{1em} \texttt{gt\_seq} & 6     & 1  / 5 & 3.41 &
    % \texttt{fold} & - \\
    % \hspace{1em} \texttt{noccs\_of\_seq} & 1     & 1  / 2 & 0.90 &
    % \texttt{fold} & - \\
    \hspace{1em} \texttt{gt\_seq} & 8     & 1  / 2 & 16.01 &
    \texttt{filter} & - \\
    \hspace{1em} \texttt{counter\_filter\_seq} & 1     & 1  / 2 & 0.32 &
    \texttt{filter} & \check \\
    \hspace{1em} \texttt{counter\_map\_seq} & 1     & 1  / 2 & 0.29 &
    \texttt{map} & \check \\
    \rowcolor{thegray}
    Graphs & 275    & 124 / 286 &  &  &  \\
    \hspace{1em} \texttt{intersect} & 13     & 12  / 15 & 3.13  &
    \texttt{fold + fold} & - \\
    \hspace{1em} \texttt{union} & 11     & 4  / 17 & 4.40 &
    \texttt{fold + fold} & - \\
    \hspace{1em} \texttt{complement} & 12     & 8  / 23 & 5.80  &
    \texttt{fold + fold} & - \\
    \hspace{1em} \texttt{mirror} & 6     & 5  / 8 & 2.51  &
    \texttt{fold} & - \\
    \hspace{1em} \texttt{copy\_vertices} & 1     & 1  / 2 & 1.64  &
    \texttt{fold} & - \\
    \hspace{1em} \texttt{check\_path} & 71     & 44  / 103 & 47.79  &
    \texttt{iter} & \check \\
    \hspace{1em} \texttt{check\_path$^{\dagger}$} & 161     & 50  / 118 & 105.07  &
    - & \check \\
    %% \hspace{1em} \texttt{check\_path\_dc} & 213     & 50  / 118 & 108.66  & - & - \\
    \rowcolor{thegray}
    Binary trees& 14     & 5  / 6  &  & & \\
    \hspace{1em} \texttt{sum\_tree} & 5     & 1  / 2 & 10.51  &
    \texttt{fold} & - \\
    \hspace{1em} \texttt{height\_tree} & 1  & 1  / 2 & 1.02  &\texttt{fold} & - \\
    \hspace{1em} \texttt{gt\_tree} & 8 & 3  / 2 & 37.05  &\texttt{iter} & \check \\ \hline
  \end{tabular}
  \caption{Summary of verified iteration case studies.}
  \label{tab:summary}
\end{table}

\par \emph{Sequences.}
The \gosp{sum_seq} function generates more VCs than others in the
module, due to \gosp{split_vc} and \gosp{in_line} goals introduced by
a \whythree strategy. Yet, examples with a single VC suggest
this splitting was unnecessary. Although \gosp{queue_of_seq} and
\gosp{stack_of_seq} are similar, the latter's specification includes a
sequence reversal, adding a slight overhead to proof replay. The
\gosp{gt_seq} function counts elements above a threshold by filtering
and then measuring the length of the resulting sequence. Finally,
\gosp{counter_filter_seq} and \gosp{counter_map_seq} are simple
iterators: one filters positives, the other increments elements. Both
write to a reference, but their effects are correctly identified
by the consumer functions.

% Overall, provers have some difficulty handling the recursive function
% \gosp{sum}, contributing to the slower verification of \gosp{sum_seq}.

\par \emph{Graphs.}
All graph operations are fully automatic. The slowest among these is
the complement of a graph, essentially due to its verbosity and need
to keep track of the universe of vertices that have not been
explored. This verbosity is observable when relating the number of
lines of specification (LoS) of each case study to their proof time,
which tends to grow linearly. Within the \gosp{Check}\footnotemark{}
module, we revisited a path-checking algorithm, \gosp{check_path}. A
version of this function, proved correctly in previous
work~\cite{castanho}, \gosp{check_path}$^{\dagger}$, relied on
manually deconstructing the iteration into a recursive function and
then specifying the iteration. In contrast, our approach allows for a
direct specification of the higher-order iterators, resulting in half
as many VCs, thus, halving the proof time.

\footnotetext{\url{https://github.com/backtracking/ocamlgraph/blob/master/src/path.ml}}

\par \emph{Binary Trees.}  We specified iteration over binary trees
using classical structures. The \gosp{permitted} predicate flattens a
tree into a sequence and reasoning about it as such. The sum of
integers values of a tree \gosp{sum_tree} has similar performance as
its counterpart \gosp{sum_seq}, though facing similar challenges. For
computing tree height, we used a different higher-order iterator,
\gosp{fold_level}, which proceeds level by level, akin to
breadth-first search. Its \gosp{permitted} predicate ensures that each
position \emph{i} of the visited sequence holds all elements at level
\emph{i}. The height then follows directly from this
specification. Alternatively, in \gosp{gt_tree}, we used an
\gosp{iter} to count elements exceeding a threshold. The impact in
performance is clearly noticeable, this can be partially due to either
the fact that we are flatting the tree into a sequence or using an
iterator with effects.

\par In any case, we observe that the number of lines of specification
(LoS) tends to exceed the number of lines of code (LoC). This is not
surprising, as the specification effort is generally greater than the
implementation. %  The use of higher-order iteration allows for
% more modular and concise implementations, but at the cost of more
% elaborate specifications.
%
The benchmarks were conducted on a machine with a processor i5-2520M
2-cores @ 3.2GHz; 12Gb RAM; 64-bit Linux Kernel 6.9.5. The proof replay
times are results of 5 executions of \whythree's \texttt{replay}
command, using \gosp{hyperfine} with 3 warm-up executions.

\section{Related work}

Recently, several works have been published addressing the formal
verification of higher-order iterators. Like our approach, these works
are based on a specification methodology that exploit the pair of
predicates \of{permitted} and \of{complete}. However, the works
described below differ from our presentation primarily in their target
language (none use \ocaml) and the verification toolchain employed.

Regarding the \textsf{Rust} programming language, we find the works of
Denis and Jourdan~\cite{denis2023}, as well as the work by Bílý
\textit{et al.}~\cite{bily}. The former use the \of{permitted} and
\of{complete} predicates to formally verify higher-order iterators
using the \of{Creusot} tool~\cite{DBLP:conf/icfem/DenisJM22}. The
latter adopt the same underlying methodology but carry out the
verification process in
\of{Prusti}~\cite{DBLP:journals/pacmpl/Astrauskas0PS19}, a tool that
translates \textsf{Rust} programs into the \of{Viper} verification
language~\cite{viper}. Unlike traditional functional languages, \textsf{Rust}
does not provide a type for anonymous functions. It is up to the user
to decide, in each case, the appropriate kind of closure to use:
\gosp{Fn}, \gosp{FnMut}, or \gosp{FnOnce}, that is, by ownership,
mutable reference, or immutable reference. This decision has a direct
impact on the verbosity of the resulting specification. In \ocaml, and
in our approach in particular, we support any consumer function,
regardless of its nature, without affecting the generated
specification.

Additionally, Pottier's work~\cite{hashtable} represents a
significant application of the \of{permitted}$/$\of{complete}
methodology outside the scope of an automatic verification
platform. In this work, the author uses the \cfmllang~\cite{cfml} tool
to formally verify the implementation of a hash table from the \ocaml
standard library, as well as the iterators provided for this data
structure. The proofs are carried out entirely in \coq, which demands
a high level of proof effort and human interaction, especially when
compared to our approach using \cameleer.

\section{Conclusions \& future work}

We developed an extension to \gospellang focused on the specification
of higher-order iterators. We also developed an extension to \cameleer
that could translate these specifications into regular \whyml code
that could, in turn, be deductively verified using \whythree.
Additionally, we devised a collection of case studies that
showcase how one might use our work to verify
higher-order iterators, through the \gosp{permitted} and
\gosp{complete} predicates.
We tackled the verification
of realistic \ocaml code, as taken from the \ocamlgraph library, which
shows that our methodology scales-up well in practice.  In summary, we
presented a simple, generic and modular specification to capture
higher-order iteration. This specifications abstracts away how the
iteration is implemented, focusing only on the important
logical elements of an iteration: the \gosp{permitted}/\gosp{complete}
relations and a user-supplied invariant.

In the following, we highlight some lines of possible future work.

\par \emph{Generic specification for patterns.} Our generic iteration
specification captures patterns that any client or iterator must
follow, regardless of their concrete implementation. % We believe this
% pattern-based specification could be applied to other common \ocaml
% language functions, such as the higher-order functions \gosp{map} or
% \gosp{filter}.
This specification would introduce a concept similar to
\emph{typeclasses} or \emph{traits} in \gospellang, as it would only
be necessary to describe the general abstractions and provide concrete
instances for each specific use.

\par \emph{Relational equivalence.} The correctness of our translation
schemas is based on an informal argument that a higher-order iterator
can be converted in a cursor that plays the iteration part. One
possible extension to our methodology could be to generate a skeleton
of a proof, for example, in \coq that captures such equivalence. It
would be natural to resort to binary logic, \emph{e.g.}, Relational
Hoare Logic~\cite{benton}, allowing us to reason about the equivalence
of two \ocaml programs.

\par \emph{Verification of Iteration Implementations.}
While we focused on specifying higher-order iterators and translating
them into cursor-based definitions and clients, we have yet to verify
existing implementations against these specifications. In that regard,
we propose verifying that higher-order function implementations
conform to their \gosp{permitted}/\gosp{complete} based
specifications, likely requiring formal proofs in frameworks like
\cfmllang or \textsf{Iris}~\cite{jung2018iris}.

% \subsubsection{Acknowledgments.}
% This work is partially financed by Agence Nationale de la Recherche
% (ANR) project ANR-22-CE48-0013-01 (GOSPEL).

\bibliographystyle{splncs04}
\bibliography{bibliography}

\end{document}

%% file: config.tex
\usepackage[T1]{fontenc}
\usepackage{graphicx}
\usepackage{subcaption}
\usepackage{xspace}
\usepackage{url}
\usepackage{textcomp}

\usepackage{amsmath,amssymb,latexsym}
\usepackage{mathabx}

\usepackage{hyperref}

\usepackage{cleveref}
\usepackage{listings}
\usepackage{xcolor}
\usepackage{bold-extra}
\usepackage{listings}
\usepackage{tikz}
\usepackage[skins]{tcolorbox}
\usepackage{microtype}
\usepackage{cellspace}

\usepackage{color, colortbl}

\usepackage{array}

\usepackage[os=win, mackeys=symbols]{menukeys}

\renewcommand{\check}{
  \raisebox{-.25\height}{\includegraphics[width=1em,height=1em]{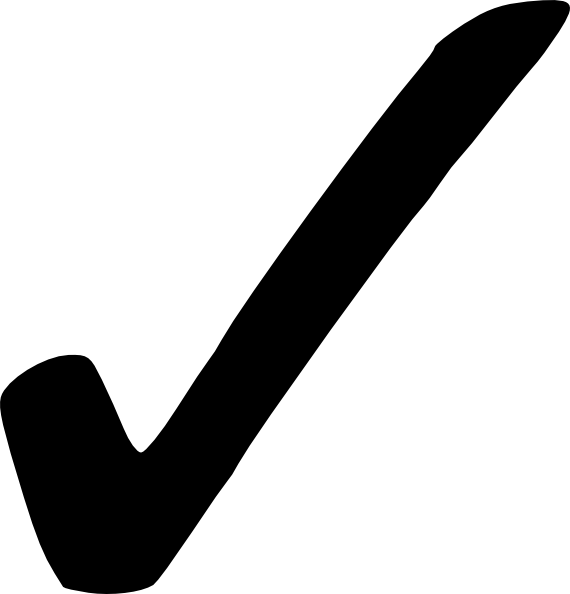}}}

\definecolor{thegray}{rgb}{0.849,0.849,0.849}

\definecolor{codemarkcolor}{RGB}{198, 210, 201} % Define the color (you can adjust the RGB values)
\newcommand{\codemark}[1]{%
  \hspace{4pt}% Adjust the spacing here
  \tikz[overlay,remember picture,baseline=(#1.base)]\node[draw,circle,inner sep=0.7pt,yscale=0.8,xscale=0.8,fill=codemarkcolor] (#1) {\texttt{#1}};
  \hspace{6pt}% Adjust the spacing here
}
\newcommand{\codemarkt}[1]{%
  \hspace{4pt}% Adjust the spacing here
  \tikz[overlay,remember picture,baseline=(#1.base)]\node[draw,circle,inner sep=0.7pt,yscale=0.8,xscale=0.8,fill=codemarkcolor] (#1) {\texttt{#1}};
  \hspace{1pt}% Adjust the spacing here
}

\usepackage{gospel}
\DeclareCaptionFormat{listing}{\hfill#3}
\usepackage{stackengine}
\def\defen{\triangleq}

%% file: mymacros.tex
\usepackage{color}

\newcommand{\whyml}{\textsf{WhyML}\xspace}

\newcommand{\ocaml}{\textsf{OCaml}\xspace}

\newcommand{\coq}{\textsf{Rocq}\xspace}
\newcommand{\isabelle}{\textsf{Isabelle}\xspace}

\newcommand{\cfmllang}{\textsf{CFML}\xspace}

\newcommand{\gospellang}{\textsf{Gospel}\xspace}
\newcommand{\cameleer}{\textsf{Cameleer}\xspace}
\newcommand{\cvc}{\textsf{CVC5}\xspace}
\newcommand{\altergo}{\textsf{Alt-Ergo}\xspace}
\newcommand{\zthree}{\textsf{Z3}\xspace}
\newcommand{\whythree}{\textsf{Why3}\xspace}
\newcommand{\ocamlgraph}{\textsf{OCamlGraph}\xspace}

\definecolor{thegray}{rgb}{0.9,0.9,0.9}
\definecolor{colorspec}{rgb}{0,0,0.797}
\definecolor{thered}{rgb}{0.797,0,0}
\definecolor{darkgreen}{rgb}{0.797,0,0}
\definecolor{theblue}{rgb}{0,0,0.797}
\definecolor{darkgray}{rgb}{0.8477,0.8477,0.8477}
\definecolor{ocaml-bg}{rgb}{0.9,0.9,0.9}
\definecolor{thegraygray}{rgb}{0.5,0.5,0.5}